\documentclass[twocolumn]{aastex631}	
\graphicspath{{./}{figures/}}
\usepackage{multirow}
\usepackage{float}
\usepackage{hyperref}
\usepackage{mathtools}
\usepackage{comment}

\begin{document}

\title{X-ray Properties of Optically Variable Low-mass AGN Candidates}

\author{Alexander Messick}
\affiliation{Washington State University, Department of Physics and Astronomy, WA 99164, USA}

\author{Vivienne Baldassare}
\affiliation{Washington State University, Department of Physics and Astronomy, WA 99164, USA}

\author{Marla Geha}
\affiliation{Yale University, Department of Astronomy, New Haven, CT 06511, USA}

\author{Jenny Greene}
\affiliation{Princeton University, Department of Astrophysical Sciences, Princeton, NJ 08544, USA}

\begin{abstract}
We present an X-ray analysis of fourteen nearby ($z<0.044$) AGN in low mass galaxies ($M_*\lesssim5\times10^{9}M_\odot$) selected based on their optical variability \citep{2020ApJ...896...10B}. 
Comparing and contrasting different AGN selection techniques in low-mass galaxies is essential for obtaining an accurate estimate of the active fraction in this regime.
We use both new and archival observations from the \textit{Chandra X-ray Observatory} to search for X-ray point sources consistent with AGN.
Four objects have detected nuclear X-ray emission with luminosities ranging from $L_{0.5-7}\approx3\times10^{40}$ to $9\times10^{42} \rm{erg\,s^{-1}}$ with two more marginal detections.
All of the detected galaxies have luminosities exceeding those anticipated from X-ray binaries, and all sources are nuclear, suggesting the X-ray emission in most sources is due to an AGN.
These observations demonstrate the success of variability at identifying AGN in low-mass galaxies.
We also explore emission line diagnostics and discuss the differences in the results of these methods for AGN selection, in particular regarding low-mass and low-metallicity systems.
\end{abstract}

\section{Introduction} \label{sec:intro}
Supermassive black holes (BHs; $M_{BH}\gtrsim 10^5 M_\odot$) occupy the centers of most massive ($M_*\gtrsim 10^{10} M_\odot$) galaxies, including our own \citep{2020IAUS..353..186K}.
Less is known, however, about their prevalence in dwarf galaxies ($M_*\lesssim 10^{10} M_\odot$).
This information could prove vital toward understanding the formation of massive BHs.
In particular, the occupation fraction of BHs in the centers of dwarf galaxies could place constraints on BH seed formation mechanisms by providing insight into the role of BHs in the evolution of their host galaxies \citep{2020ARA&A..58..257G, 2021NatRP...3..732V}.

Historically, detecting nucleic BHs has proven to be difficult since their gravitational spheres of influence are typically too small to be dynamically resolved in galaxies beyond $\sim5$ Mpc.
For this reason, much of the search for massive BHs in low-mass galaxies focuses on signs of accretion as evidence for the presence of an active galactic nucleus (AGN).
Large-scale spectroscopic surveys have allowed for the detection of AGN signatures using methods such as narrow-line diagnostics, which use ratios of narrow-line emission fluxes to classify galaxies \citep[BPT diagrams;][]{Baldwin_1981, 2006MNRAS.372..961K}.
While these methods have been successful in identifying a large number of new AGN \citep{2013ApJ...775..116R, 2014AJ....148..136M}, there likely exists a population that is undetected by optical spectroscopy due to selection effects \citep{2015ApJ...811...26T, 2018ApJ...861...50B}.
In particular, low-mass and low-metallicity systems tend to have decreased [NII]-to-H$\alpha$ ratios, pushing them to the left and into the star-forming regime on the BPT diagram \citep{2003MNRAS.346.1055K, Cann_2019}.
Using alternate AGN selection techniques is necessary for obtaining an accurate estimate of the active fraction in dwarf galaxies (and thus constraining the occupation fraction).

Long-term optical variability has been a prolific tool for identifying bright AGNs \citep{doi:10.1146/annurev.astro.35.1.445, 2003AJ....125....1G, Schmidt_2010}.
More recently this has been used to identify AGN in low-mass or low-metallicity galaxies \citep{2018ApJ...868..152B,2020ApJ...896...10B,2021MNRAS.504..543B,2022ApJ...930..110Y}.
\citet{2020ApJ...896...10B} used data from the Palomar Transient Factory (PTF) to search for variations in the light outputs for 35,000 low-mass galaxies, finding 238 objects in this low-mass regime.
Of this population, 75\% had narrow emission lines dominated by star formation and would have been missed by performing BPT analysis alone.

X-ray observations can provide confirmation of the presence of an AGN and insight into its accretion properties.
Although some work has been done on X-rays from low-mass AGN \citep{2009ApJ...698.1515D,2012ApJ...761...73D}, most of the research on the topic has focused on galaxies more massive than dwarfs.
Additionally, X-ray emission from low-mass AGN can be difficult to disentangle from emission from X-ray binaries (XRBs).
Nevertheless, X-ray observations are thought to be one of the most reliable techniques for AGN selection, since X-ray emission from non-AGN astrophysical processes is typically comparatively weak \citep{2018ARA&A..56..625H}.
In this work, we explore the X-ray properties of low-mass galaxies with AGN-like optical variability.
Our goals are to contrast different AGN selection techniques at the low-mass end and understand selection biases.

This paper is organized as follows: In Section \ref{sec:data}, we discuss the selection of the original sample of galaxies and the origin for all the data.
In Section \ref{sec:analysis}, we introduce the sources for all the data and the methods of analysis, displaying the results of the calculation in Section \ref{sec:results}.
We then discuss the implications of the results in Section \ref{sec:discussion}, comparing the outcomes of the different selection methods and calculating the Eddington luminosities and ratios of the objects in the sample.

\begin{deluxetable*}{c c c c c c c c c}[htb]
\tablecolumns{8}
\tablecaption{Variability-Selected AGN with X-ray Follow-Up \label{table:pos}}
\tablehead{
    \colhead{NSA ID} & \colhead{R.A.} & \colhead{Decl.} & \colhead{Radius} & \colhead{Redshift} & \colhead{Distance} & \colhead{$\log{\mathrm{M_*}}$} & \colhead{BPT Class} & \colhead{FeX Emission}\\
    & (hms) & (dms) & (arcsec.) & (z) & (Mpc) & ($\log{\mathrm{M_\odot}}$) & & \\
    (1) & (2) & (3) & (4) & (5) & (6) & (7) & (8) & (9)
    }
\startdata
    10045  & 08:29:12.67 & +50:06:52.8 & 3.072  & 0.043 & 186.3 & 9.65 & AGN   & y \\
    104881 & 08:00:28.55 & +15:27:11.2 & 7.675  & 0.015 & 68.7  & 9.04 & SF    & n \\
    124477 & 12:21:34.09 & +04:46:46.3 & 25.354 & 0.007 & 31.4  & 9.67 & ---   & n \\
    124554 & 11:40:40.73 & +59:48:50.3 & 8.479  & 0.012 & 50.2  & 9.09 & ---   & n \\
    15235  & 14:40:12.70 & +02:47:43.5 & 7.590  & 0.030 & 126.6 & 9.49 & AGN   & y \\
    152627 & 23:47:04.69 & +29:28:56.2 & 43.790 & 0.017 & 75.1  & 9.48 & Comp. & n \\
    156688 & 08:27:23.93 & +23:10:48.4 & 16.859 & 0.018 & 78.4  & 7.66 & Comp. & n \\
    51928  & 09:44:19.41 & +09:59:05.3 & 8.752  & 0.010 & 45.3  & 9.41 & SF    & n \\
    57867  & 16:19:02.49 & +29:10:22.2 & 6.985  & 0.009 & 36.6  & 7.69 & SF    & n \\
    61072  & 12:39:59.29 & +47:38:49.7 & 7.086  & 0.031 & 134.9 & 9.11 & SF    & n \\
    67333  & 12:26:47.95 & +07:40:17.6 & 11.924 & 0.002 & 11.3  & 8.24 & ---   & n \\
    86652  & 12:42:06.47 & +33:16:43.9 & 8.978  & 0.038 & 164.2 & 9.40 & SF    & n \\
    88260  & 12:38:56.92 & +38:05:24.9 & 9.449  & 0.007 & 33.7  & 8.45 & SF    & n \\
    97904  & 15:39:50.62 & +21:43:22.8 & 3.857  & 0.038 & 164.3 & 9.61 & ---   & n \\
\enddata
\tablecomments{Col. (1) gives each object's ID in the NASA-Sloan Atlas. Cols. (2) and (3) give the Right Ascension and declination in units of hours:minutes:seconds and degrees:arcminutes:arcseconds, respectively. Col. (4) gives the Petrosian 90\%-light radius in arcseconds, while Cols. (5), (6), and (7) give each object's redshift, distance in megaparsecs, and mass in solar masses. We assume $h=0.7$. Col. (8) gives the preliminary class found from BPT analysis \citep{2020ApJ...896...10B}, and Col. (9) gives the whether the \ion{Fe}{10} coronal line was found in each galaxy's spectrum.}
\end{deluxetable*}

\section{Data} \label{sec:data}
\subsection{Sample Selection} \label{subsec:selection}
Our sample consists of fourteen dwarf galaxies selected for their optical variability from \citet{2020ApJ...896...10B}, which used data from the Palomar Transient Factory (PTF) and NASA Sloan Atlas (NSA).
The original sample of galaxies were selected from the NSA for their mass ($M_*\lesssim 2\times10^{10} M_\odot$) and PTF coverage, then analyzed using difference imaging techniques to detect photometric variability consistent with AGNs \citep{doi:10.1146/annurev.astro.35.1.445, 10.3389/fspas.2017.00035}.
Since the focus of this paper is variable AGNs in low-mass galaxies, here we analyze observations of galaxies with stellar masses less than $\sim5\times10^9$ solar masses. 
These criteria resulted in a sample of 238 dwarf galaxies with AGN-like variability.
Of these galaxies, we requested new observations for eight from the Chandra Observatory, targeting the most nearby.
We were also able to find archival data for an additional six galaxies, giving our total sample of fourteen.

The NASA Sloan Atlas is a catalog of reprocessed data from the Sloan Digital Sky Survey's \citep[SDSS DR8;][]{2000AJ....120.1579Y, 2011ApJS..193...29A} five band imaging combined with GALEX's imaging in the ultraviolet.
The SDSS images are produced using sky-subtraction, detection, and deblending techniques described in \citet{2011AJ....142...31B}.
We use the catalog nsa\_v0\_1\_2.fits\footnote{\url{http://sdss.physics.nyu.edu/mblanton/v0/nsa_v0_1_2.fits}}, which corresponds to SDSS Data Release 8 and contains several relevant quantities. This version of the catalog extends out to redshift $z=0.055$.
These include positional information such as Right Ascension and Declination, as well as heliocentric redshift and a distance estimate using peculiar velocity, stellar mass from a K-correction fit (given in $M_\odot/h^{-2}$, so we assume $h=0.7$), the Petrosian 90\%-light radius derived from the r-band with corresponding axis ratio and angle, and H$\alpha$ flux for all but 2 objects, with errors reported for the distance, mass, and flux.
Some of these values are shown in Table \ref{table:pos}.

\begin{figure*}[htb]
 \centering
 \includegraphics[width=.24\linewidth]{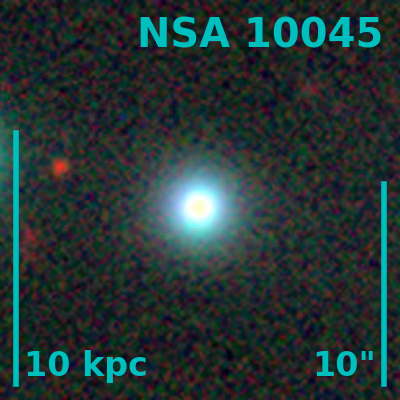} 
 \includegraphics[width=.24\linewidth]{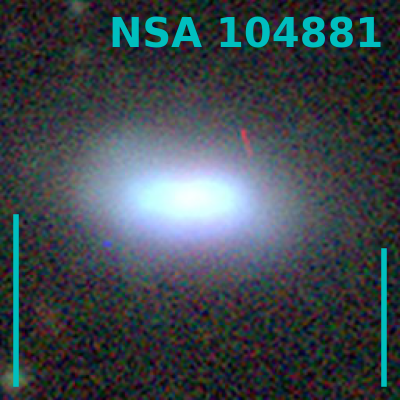} 
 \includegraphics[width=.24\linewidth]{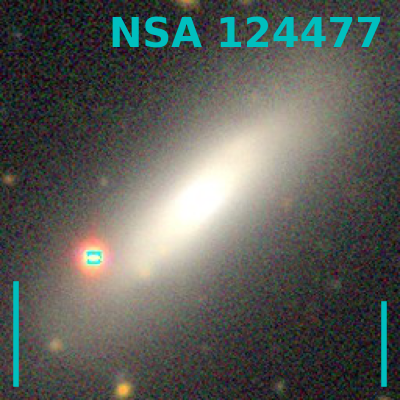} 
 \includegraphics[width=.24\linewidth]{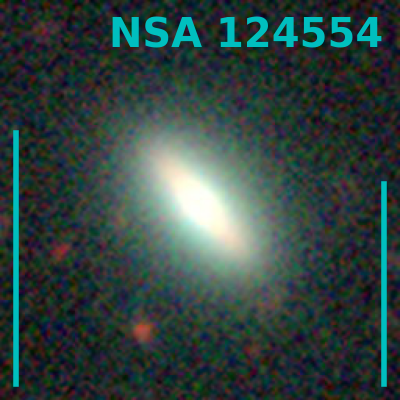} \\
 
 \includegraphics[width=.24\linewidth]{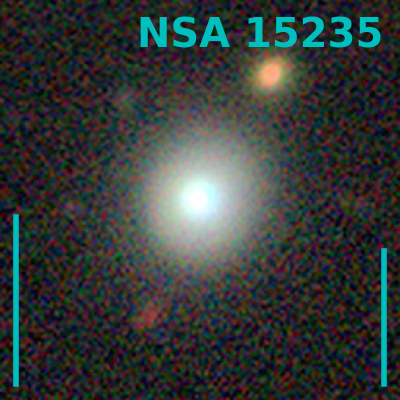} 
 \includegraphics[width=.24\linewidth]{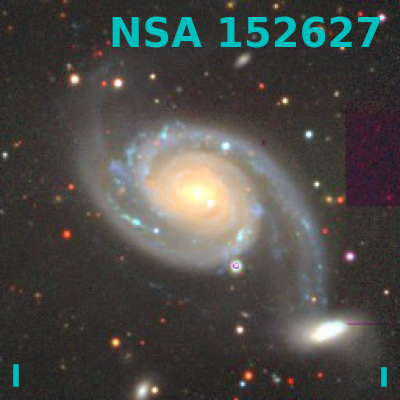} 
 \includegraphics[width=.24\linewidth]{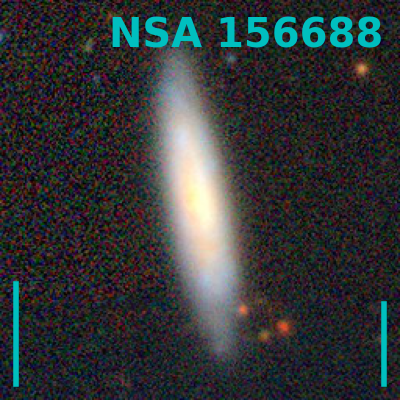} \\
 
 \includegraphics[width=.24\linewidth]{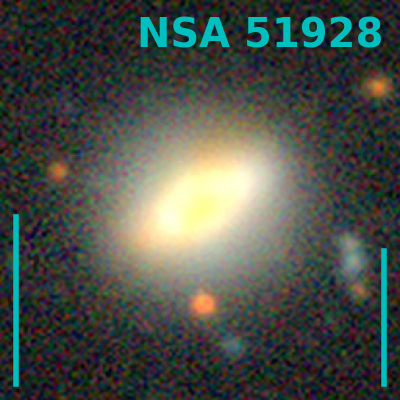}
 \includegraphics[width=.24\linewidth]{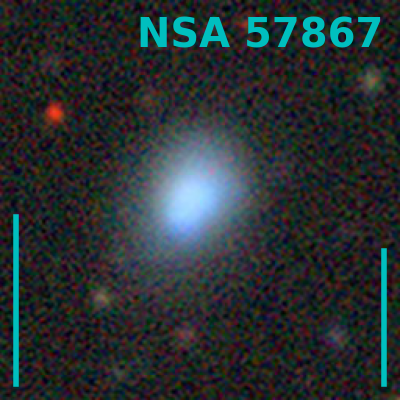}
 \includegraphics[width=.24\linewidth]{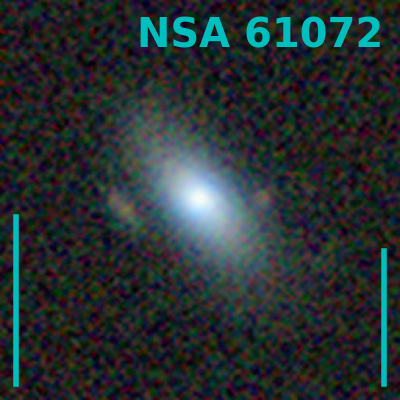}
 \includegraphics[width=.24\linewidth]{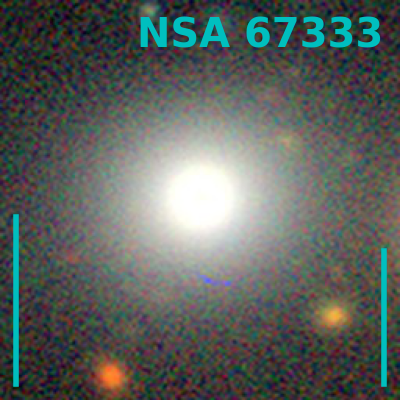} \\
 
 \includegraphics[width=.24\linewidth]{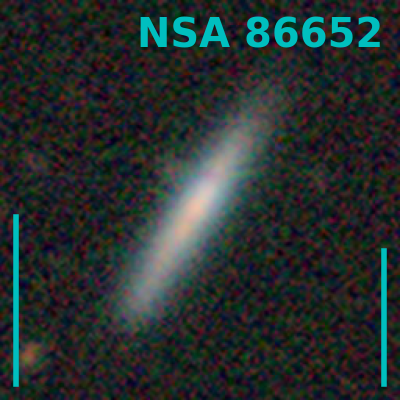}
 \includegraphics[width=.24\linewidth]{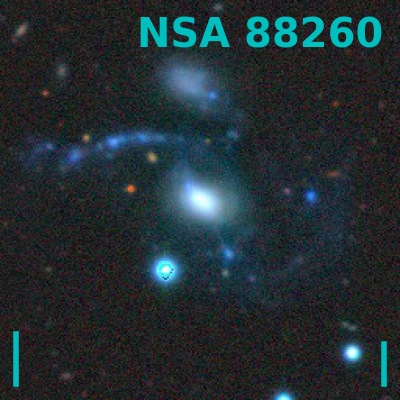}
 \includegraphics[width=.24\linewidth]{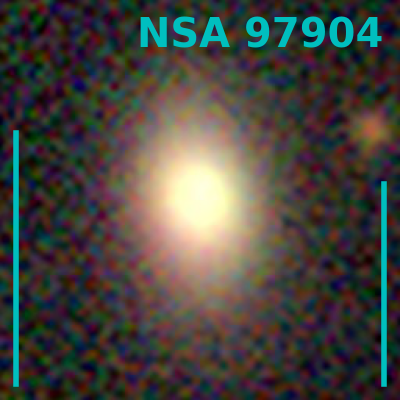} \\
 \caption{Images of the 14 objects taken from the DESI Legacy Survey \citep{2019AJ....157..168D}. Vertical bars showing 10 kiloparsecs and 10 arcseconds are shown respectively at the bottom left and right corners of each image for scale.}
\end{figure*}

\subsection{X-Ray Observations} \label{subsec:xray}
NASA's \textit{Chandra} X-ray Observatory is a space telescope that launched in 1999 designed to detect X-ray emission from hot regions of the universe.
Observations for these objects were made with the \textit{Chandra}'s Advanced CCD Imaging Spectrometer (ACIS) detector, which allows for resolution of about 1 arcsec.
Eight objects were targeted through GO 22700333 (PI: Baldassare).
These objects were chosen to be the most nearby galaxies found to have AGN-like variability in \cite{2020ApJ...896...10B}.
An additional six objects had archival \textit{Chandra} data, originally observed based on their identification as as an ultra-luminous X-ray source candidate (NSA 152627), as having peculiar velocities (NSA 97904), or as AGN candidates based on optical variability (NSA 15235), transient behavior (NSA 86652), spectroscopy (NSA 61072), or merger history (NSA 10045).
Overall, total observation times range from $\sim$1.1 to $\sim$46.8 kiloseconds.


\subsection{Ancillary Observations} \label{subsec:ancillary}
We make use of far UV (FUV; 1350-1750 \AA) observations from the Galaxy Evolution Explorer \citep[GALEX;][]{Martin_2005, 2014yCat.2335....0B}, an ultraviolet space telescope.
We use this data for NSA 152627, for which the H$\alpha$ flux is unavailable in the NSA catalog.
Data in the infrared was taken from the Wide-field Infrared Survey Explorer \citep[WISE;][]{2010AJ....140.1868W}, which gives the magnitudes of each band; we used the W4 band to calculate the 25 $\mu$m flux necessary for dust corrections.
These data are available at \cite{https://doi.org/10.26131/irsa1}.
We use single-fiber spectra from the SDSS Data Release 16 \citep[DR16;][]{2020ApJS..249....3A} to calculate the flux from [Fe X]$\lambda6374$ emission.

\section{Analysis}\label{sec:analysis}
\subsection{X-Ray Data} \label{subsec:Chandra}
Data from the Chandra Observatory is processed using the Chandra Interactive Analysis of Observations software (CIAO, version 4.13).
First, we reprocess the data to create a new event list with their particle backgrounds cleaned, which is useful for very faint observations.
Next, astrometric corrections are made by running CIAO's \texttt{WAVDETECT} function on the event file, which detects X-ray point sources.
We find sources in the broad band at scales of 1, 2, 4, 6, 8, 12, 16, 24, and 32 pixels with a signal threshold of $10^{-6}$ on the X-ray images.
If more than three sources are detected, they can be  cross-matched against a known source catalog, typically the SDSS Data Release 12 \citep{2011AJ....142...72E, 2015}.
The image is energy-filtered to the correct range ($0.5-10.0$ keV) and restricted to the appropriate chip.
If an object has multiple observations, they are merged together to create a single exposure-corrected image.
\texttt{WAVDETECT} also finds emission sources coinciding with four of the galaxies in our sample: NSA 10045, 104881, 15235, and 152627.
Although \texttt{WAVDETECT} did not find X-ray sources near NSA 156688 nor 51928, these galaxies have enough total X-ray photons to be included as marginal detections.

\begin{figure*}[htb]
 \centering
 \includegraphics[width=.3\linewidth]{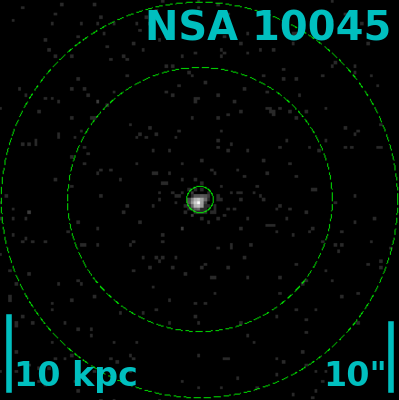} 
 \includegraphics[width=.3\linewidth]{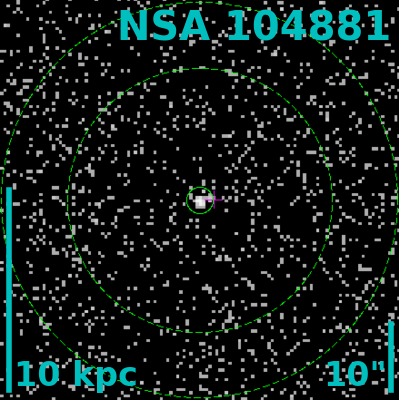} 
 \includegraphics[width=.3\linewidth]{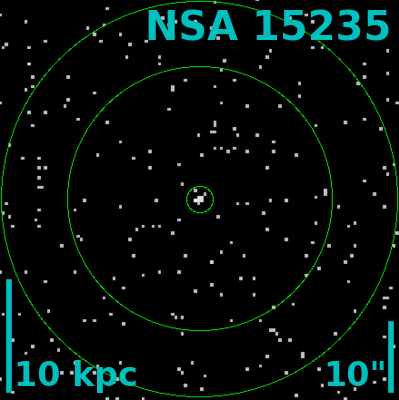}
 \includegraphics[width=.3\linewidth]{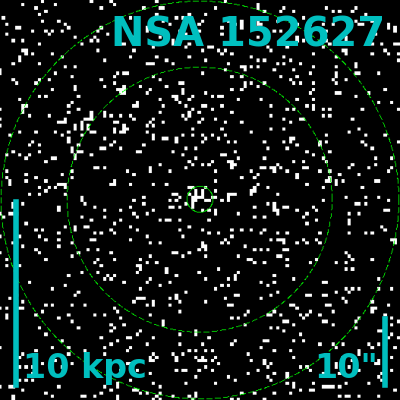}
 \includegraphics[width=.3\linewidth]{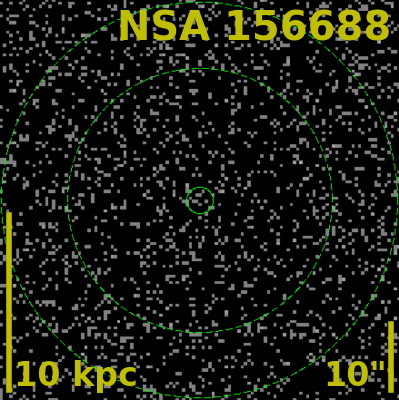}
 \includegraphics[width=.3\linewidth]{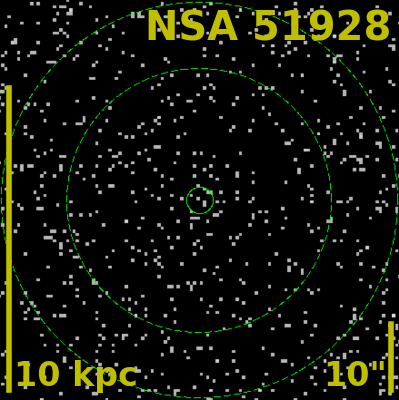} \\
 \caption{Images of the six galaxies observed in the broad X-ray band taken from the Chandra survey, with detections shown in cyan and marginal detections in yellow. Vertical bars showing 10 kiloparsecs and 10 arcseconds are shown respectively at the bottom left and right corners of each image for scale. The inner circle is the source region with a radius of 2'' and the outer two rings are the inner and outer boundaries of the annular background region with radii 20'' and 30'', respectively.}
\end{figure*}

Aperture photometry is conducted on the \textit{Chandra} data using CIAO's \texttt{SRCFLUX} software; the light signal is extracted from a $2''$ circular region around the optical galaxy center from the NSA while a source-free annulus with an inner radius of $20''$ and an outer radius of $30''$ is used for the background.
The sources are all consistent with their respective optical galaxy nuclei. 
We estimated the counts and the corresponding fluxes in the broad band ($0.5-7.0$ keV) from each object assuming an absorbed power law spectrum with a photon index $\Gamma=1.7$ with a hydrogen column in the source's direction given by the National Radio Astronomy Observatory \citep[NRAO;][]{2016A&A...594A.116H}.
Afterward, to account for contamination from X-ray binaries (XRBs), the broad band flux was converted to an unabsorbed flux in the equivalent band for the total XRB luminosity (0.5-8 keV) using the Portable Interactive Multi-Mission Simulator (PIMMS) assuming a galactic column density of $10^{21} \rm{cm^{-2}}$ and a power law source model with $\Gamma=1.8$.
The respective luminosities for all objects were then calculated from the fluxes using NSA distances.

We note that NSA 10045 had a fairly high mean count rate of 0.215 counts per second, so the effect of pileup is not negligible.
Following along the documentation provided by the \cite{pileup_abc}, we estimate the pileup fraction using:
\begin{equation}
    f=1-e^{-\Lambda_{tot}},\;
    \Lambda_{obs} = (1-f)\Lambda_{tot},
\end{equation}
where f is the total fraction of events lost from grade or energy migration and $\Lambda$ is the photon count rate in units of counts per second.
From these, we calculate a pileup fraction of roughly 10 to 25 percent, resulting in a total count rate of about 0.288 counts per second.
Unfortunately, these estimations are not precise enough to use in scientific calculations, but proper correction would only increase the observed luminosity from NSA 10045, so it does not affect the outcomes of this paper.

\subsection{Star Formation Rates and Expected Luminosity} \label{subsec:XRL}
AGNs are not the only potential X-ray sources in these galaxies; we also anticipate a populations of XRBs that could contribute to the overall X-ray luminosity ($L_X$).
\citet{2019ApJS..243....3L} demonstrates that the X-ray luminosity from low-mass and high-mass XRBs scales with star formation rate (SFR) and nuclear stellar mass ($M_*$), respectively.
They find the total expected X-ray luminosity from XRBs to be:
\begin{equation}
    \left(\frac{L_{XRB}}{erg/s}\right) =\alpha\times\left(\frac{M_*}{M_\odot}\right) + \beta\times\left(\frac{SFR}{M_\odot/yr}\right),
\end{equation}
with $\log{\alpha}=29.25$ and $\log{\beta}=39.71$.

For the stellar masses, we use the values reported in the NSA catalog.
To calculate SFR, we use the following relation given in \citet{Kennicutt_2012} between the SFR and dust-corrected luminosity of certain ``tracer'' bands:
\begin{equation}
log{\left(\frac{SFR}{M_\odot/yr}\right)} = \log{\left(\frac{L_x}{erg/s}\right)} - \log{C_x},
\end{equation}
where $C_x$ is a unitless calibration constant specific to the band $x$.
We use H$\alpha$ fluxes given by the NSA catalog for all but NSA 152627, for which H$\alpha$ data is unavailable.
For this object, we obtain GALEX observations in the FUV band.
The corresponding constants are $\log{C_{H\alpha}}=41.27$ and  $\log{C_{FUV}}=43.35$.
Aperture photometry is conducted on the GALEX data; we calculate the count rate from a circle with a radius of two arcseconds.
These count rates are then converted into flux densities according to GALEX's documentation, which are then converted into fluxes by multiplying them by 1516 \AA.
Just as with the Chandra data, these fluxes are also converted to luminosities using the distances from the NSA.

We implement dust corrections with the magnitudes reported in WISE's W4 band \citep{2010AJ....140.1868W}.
We calculate the flux density and luminosity at 25 $\mu m$ with NSA's distances, assuming the constant power law spectrum given in their documentation.
The corrections are then applied following the results of \citet{2009ApJ...703.1672K} and \citet{2011ApJ...741..124H} for FUV and H$\alpha$, respectively.

Once we have an estimate for the XRB luminosity, we must find the luminosity expected to come from the central 2''.
To do this, we assume that the distribution of XRBs follow that of the optical light within the galaxy.
We then multiply the total XRB luminosity by the ratio of each object's optical luminosity within a 2$''$ aperture to that within its 90\%-light radius, which was found in the NSA catalog.
We note that there is scatter in scaling relations used to estimate XRB luminosity, so any individual object could have an over- or under-estimated expected XRB luminosity.
Additionally, there could be contributions to the FUV or H$\alpha$ luminosity from an AGN, leading to an over-estimate of the SFR and thus of the expected XRB luminosity.

\subsection{FeX Emission}
All spectra were analyzed with a pipeline using the Python package \texttt{astropy} \citep{2013A&A...558A..33A, 2018AJ....156..123T} and its affiliate \texttt{specutils} \citep{2019ascl.soft02012A}.
First, we select for objects whose [\ion{O}{1}] flux signal-to-noise ratio is at least 3 according to the NSA file.
Then, for each spectrum we define a spectral region spanning 150 \AA around the [\ion{O}{1}]+[\ion{Fe}{10}] complex.
We fit the continuum within this region to a second order polynomial, masking small regions ($\mathrm{\sim5\AA}$ each) around the [\ion{O}{1}]$\lambda6300$, [\ion{S}{3}]$\lambda6313$ and [\ion{Fe}{10}]$\lambda6374$ lines to ensure a good fit.
The continuum is then subtracted and we fit a single Gaussian within the small region surrounding the [\ion{O}{1}]$\lambda6300$ line.
From this fitting, we define the [\ion{O}{1}]$\lambda6363$ line to have the same velocity as [\ion{O}{1}]$\lambda6300$ with a flux ratio of [\ion{O}{1}]$\lambda6300$/[\ion{O}{1}]$\lambda6363$=3.
We subtract the combined [\ion{O}{1}] doublet from each of the remaining spectra, and we fit a Gaussian within the small region surrounding the [\ion{Fe}{10}] line.
The total flux from the emission line is then the area underneath this Gaussian, which is the product of its amplitude, standard deviation, and a factor of $\sqrt{2\pi}$.
For error analysis, we resample the spectrum by adding a random error at each point drawn from a normal distribution whose scale is defined by the uncertainties reported by SDSS.
We utilize bootstrapping and repeat this process one thousand times to output a range of fluxes.
We plot a histogram to this output set, using Scott's normal reference rule to set bin widths.
For the fluxes, we set a cutoff of $10^{-20} \rm{erg\,s^{-1}\,cm^{-2}}$, above which fluxes are counted as "nonzero".
If at least 75\% of the calculated fluxes from an observation were non-zero, we count this as a detection.
For these detections, we discard the zeros and fit the remaining histogram to a Gaussian, the mean and standard deviation of which represent the final [\ion{Fe}{10}] flux and its uncertainty.

\section{Results} \label{sec:results}
\subsection{X-Ray Properties} \label{subsec:xprops}
Of the fourteen galaxies observed, four yielded detections in the broad X-ray band with luminosities ranging from $3\times10^{40}$ to $\sim9\times10^{42}\,\rm{erg\,s^{-1}}$.
For the non-detected objects, our 3-$\sigma$ upper limits range from $\sim10^{39}-2\times10^{40}\,\rm{erg\,s^{-1}}$.
The X-ray fluxes and luminosities are presented in  Table \ref{table:xray}.

\begin{figure}[htb]
    \centering
    \includegraphics[width=0.8\columnwidth]{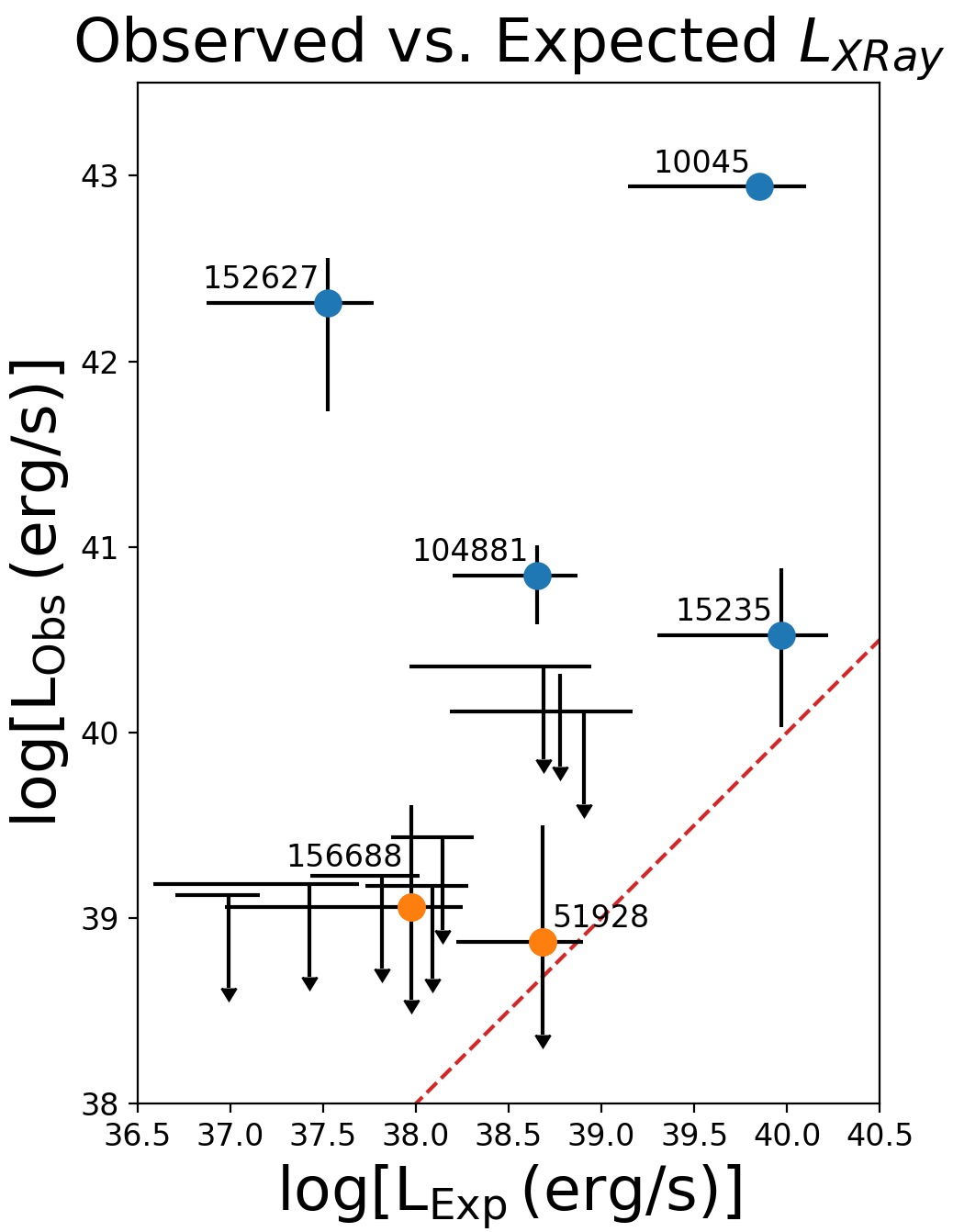}
    \caption{The logarithm of the observed X-ray luminosity versus the logarithm of the calculated expected luminosity from XRBs, with the red dashed line showing where these two values are equal. The blue points represent objects that were detected in the broad band while orange points represent marginal detections; for those that were not detected, we show their upper limits.}
    \label{fig:XLF}
\end{figure}

However, AGNs are not the only possible sources of X-rays in these galaxies.
We also expect XRBs to make a contribution to the overall X-ray luminosity of the systems.
Using the procedure described in Section \ref{subsec:XRL}, we estimate the likely X-ray binary luminosity for each object depending on SFR and stellar mass \citep{2019ApJS..243....3L}.
The estimated XRB luminosities range from $\sim 10^{37}-10^{40}\;\rm{erg\,s^{-1}}$.
All of the detected X-ray sources have luminosities exceeding the predicted XRB luminosity, while the rest are consistent.
Figure \ref{fig:XLF} shows the observed versus expected X-ray luminosities for all objects in the sample.
Table \ref{table:sfr} gives the star formation rates, and observed and predicted X-ray luminosities.
We note that if there is truly an AGN in these galaxies, there will be some contribution to both the H$\alpha$ and FUV fluxes. This would likely lead to an overestimation to the SFR and therefore the expected XRB luminosity, but this is difficult to account for. We discuss this further in Section 5.1. 


\begin{deluxetable*}{c c |CCC|CCC|CCC}[htb]
\tabletypesize{\footnotesize}
\tablecolumns{11}
\tablecaption{\label{table:xray}}
\tablehead{
    \colhead{NSA ID} & \colhead{Obs. Time} & \multicolumn{3}{c}{Counts} & \multicolumn{3}{c}{Flux ($10^{-15} \rm{erg\,s^{-1}\,cm^{-2}}$)} & \multicolumn{3}{c}{Luminosity ($\rm{log(erg\,s^{-1})}$)} \\
    \cline{3-5} \cline{6-8} \cline{9-11}
                 & (ks)  & \colhead{0.5–7 keV}  & \colhead{2-10 keV}   & \colhead{0.5-10 keV} & \colhead{0.5–7 keV}  & \colhead{0.5-8 keV}  & \colhead{2-10 keV} & \colhead{0.5–7 keV}   & \colhead{0.5-8 keV}   & \colhead{2-10 keV} \\
    (1) & (2) & (3) & (4) & (5) & (6) & (7) & (8) & (9) & (10) & (11)
    }
\startdata
    10045$^{a}$  & 5.18  & 1015.0\pm95.6        & 200.0\pm42.4         & 1017.0\pm95.7        & 1740_{-160}^{+160}   & 2100_{-200}^{+200}   & 1030_{-220}^{+220} & 42.86_{-0.04}^{+0.04} & 42.94_{-0.04}^{+0.04} & 42.63_{-0.10}^{+0.08} \\ 
    104881       & 33.76 & 44.0_{-17.3}^{+23.8} & 27.0_{-13.0}^{+19.6} & 45.0_{-17.5}^{+24.0} & 103_{-47}^{+47}      & 124_{-57}^{+57}      & 103_{-62}^{+62}    & 40.77_{-0.27}^{+0.16} & 40.85_{-0.26}^{+0.13} & 40.76_{-0.40}^{+0.21} \\    
    124477       & 7.84  & 2.0_{-1.9}^{+8.9}    & 1.0_{-1.0}^{+7.9}    & 2.0_{-1.9}^{+8.9}    & 3.5_{-3.5}^{+15.6}   & 4.3_{-4.3}^{+18.8}   & 2.1_{-2.1}^{+23.4} & 38.62^{+0.73}         & 38.70^{+0.73}         & 38.40^{+1.08}         \\
    124554       & 22.13 & \leq6.6              & \leq6.6              & \leq6.6              & \leq4.1              & \leq4.9              & \leq6.8            & \leq39.09             & \leq39.17             & \leq39.31             \\
    15235$^{a}$  & 8.03  & 10.0_{-6.9}^{+13.6}  & 1.0_{-1.0}^{+7.9}    & 10.0_{-6.9}^{+13.6}  & 14.5_{-9.9}^{+18.6}  & 17.5_{-11.9}^{+22.5} & 2.9_{-2.9}^{+24.4} & 40.44_{-0.50}^{+0.36} & 40.53_{-0.50}^{+0.36} & 39.74^{+0.98}         \\
    152627$^{a}$ & 12.15 & 17.0_{-9.8}^{+16.4}  & 2.0_{-1.9}^{+8.9}    & 17.0_{-9.8}^{+16.4}  & 2530_{-1870}^{+1880} & 3060_{-2260}^{+2260} & 578_{-578}^{+1690} & 42.23_{-0.58}^{+0.24} & 42.31_{-0.58}^{+0.24} & 41.59^{+0.59}         \\
    156688       & 46.81 & 5.0_{-4.2}^{+11.0}   & 3.0_{-2.8}^{+9.7}    & 5.0_{-4.2}^{+11.0}   & 1.3_{-1.3}^{+3.3}    & 1.6_{-1.6}^{+4.0}    & 0.6_{-0.6}^{+4.9}  & 38.98^{+0.55}         & 39.06^{+0.55}         & 38.65^{+0.96}         \\
    51928        & 16.27 & 3.0_{-2.8}^{+9.7}    & 4.0_{-3.5}^{+10.4}   & 5.0_{-4.2}^{+11.0}   & 2.5_{-2.5}^{+8.2}    & 3.0_{-3.0}^{+9.8}    & 5.3_{-5.3}^{+14.9} & 38.79^{+0.63}         & 38.87^{+0.63}         & 39.12^{+0.58}         \\
    57867        & 11.38 & \leq6.6              & \leq6.6              & \leq6.6              & \leq7.9              & \leq9.5              & \leq13.3           & \leq39.10             & \leq39.19             & \leq39.33             \\
    61072$^{a}$  & 5.18  & \leq6.6              & \leq6.6              & \leq6.6              & \leq8.6              & \leq10.4             & \leq28.8           & \leq40.27             & \leq40.35             & \leq40.80             \\
    67333        & 1.13  & \leq6.6              & \leq6.6              & \leq6.6              & \leq71.6             & \leq86.4             & \leq137            & \leq39.04             & \leq39.12             & \leq39.32             \\
    86652$^{a}$  & 13.53  & 1.0_{-1.0}^{+7.9}    & \leq6.6              & 1.0_{-1.0}^{+7.9}    & 0.5_{-0.5}^{+4.8}    & 0.6_{-0.6}^{+5.8}    & \leq13.3           & 39.20^{+1.04}         & 39.28^{+1.04}         & \leq40.63             \\
    88260        & 8.51  & \leq6.6              & 1.0_{-1.0}^{+7.9}    & 1.0_{-1.0}^{+7.9}    & \leq10.3             & \leq12.4             & 1.6_{-1.6}^{+21.6} & \leq39.15             & \leq39.23             & 38.32^{+1.17}         \\
    97904$^{a}$  & 36.80 & \leq6.6              & \leq6.6              & \leq6.6              & \leq3.3              & \leq4.0              & \leq7.8            & \leq40.03             & \leq40.12             & \leq40.40             \\
\enddata
\tablecomments{Observed X-ray data for the objects in our sample. Col. (2) gives the total Chandra observation time for each object in kiloseconds. Cols. (3), (4), and (5) give the photon counts in the broad (0.5-7 keV) and 2-10 keV bands, as well as the total X-ray photons (0.5-10 keV). 3-$\sigma$ errors are given for each value, calculated using Poisson statistics for objects with fewer than 100 counts and Gaussian statistics otherwise. Cols. (6) and (8) respectively are the fluxes corrsponding to Cols. (3) and (4), while Col. (7) is the flux in the 0.5-8 keV band calculated from the broad band using PIMMS. Cols. (9), (10), and (11) give the logarithm of the luminosities corresponding to Cols. (6), (7), and (8), respectively. Objects marked with the superscript `a' are from the \textit{Chandra} archive. Values shown with $\leq$ are upper limit values.}
\end{deluxetable*}

\subsection{Emission Line Diagnostics} \label{subsec:BPT}

The BPT diagram uses optical emission line ratios to characterize the dominant source of photoionization in a galaxy  \citep{Baldwin_1981}.
This diagram uses the ratios of optical narrow emission lines to determine the source of ionization in an object by placing the data points in one of three regions: star-forming (SF), AGN, or AGN-SF composite.

In particular, for a system whose primary source of ionization is an active nucleus, \citet{10.1111/j.1365-2966.2006.10812.x} shows the line ratio [\ion{N}{2}]$\lambda$6584 Å/H$\alpha$ to be positively correlated with both the metallicity and galaxy mass, and [\ion{O}{3}]$\lambda$5007 Å/H$\beta$ is associated with the average ionization state and temperature of the gas and decreases with the increasing contribution of star-formation to the emission-line spectrum.

\begin{figure}[htb]
    \centering
    \includegraphics[width=0.8\columnwidth]{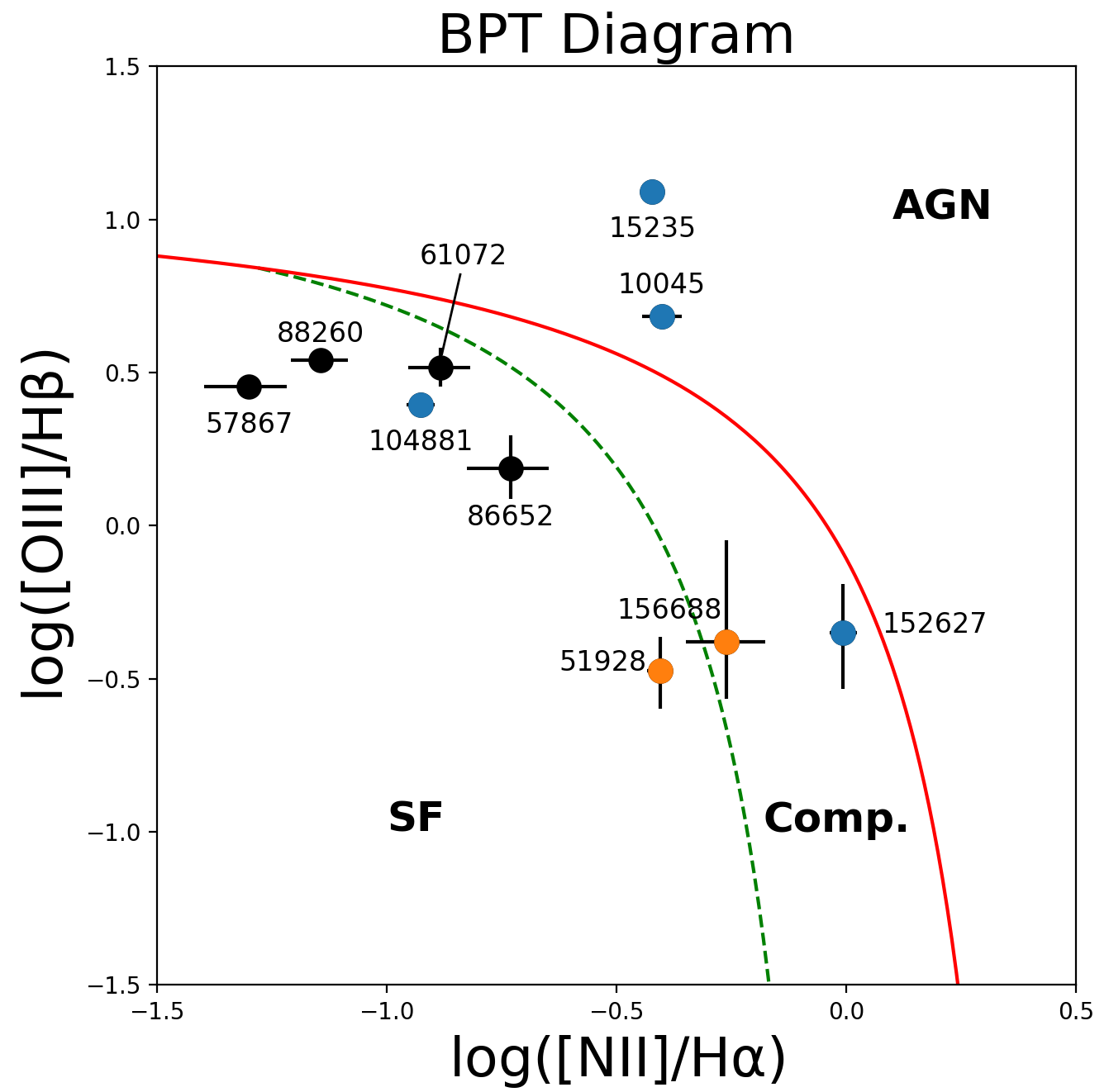}
    \caption{A BPT diagram containing the ten applicable galaxies with 3-$\sigma$ error bars, labeled by their NSA IDs. Points below the green dashed line fall within the star-forming (\textbf{SF}) region, objects above the red solid line are in the \textbf{AGN} region, and those between the two lines are in the composite (\textbf{Comp.}) region. The blue points represent objects that were detected in the broad X-ray band, orange points represent marginal detections, and black points were not detected.}
    \label{fig:BPT}
\end{figure}

We use emission lines measurements from \citet{2020ApJ...896...10B}, to place these objects on the BPT diagram when possible.
Four objects have spectra dominated by absorption for at least one of the necessary emission lines (NSA 124477, 124554, 67333, and 97904), so we do not plot these on the BPT diagram.
The remaining ten objects are plotted in Figure \ref{fig:BPT}, which categorizes six objects in the star-forming region of the diagram, two in the composite region, and two in the AGN region.

Of the X-ray detected galaxies, one is BPT star forming (NSA 104881), one is composite (NSA 152627), and two are AGN (NSA 10045 and 15235).
The marginally detected galaxies (NSA 156688 and 51928) fall in the composite and star-forming regions.
Additionally, NSA 15235 and 10045 also have broad H$\alpha$ emission lines.
The results of this analysis are also shown in Column (8) of Table \ref{table:pos}.

\begin{deluxetable}{ CC | CC }[htb]
\tablecolumns{4}
\tablecaption{\label{table:sfr}}
\tablehead{\colhead{NSA ID} & \colhead{SFR ($\rm{M_\odot\,yr^{-1}}$)} & \multicolumn{2}{c}{X-Ray Luminosity ($\rm{log(erg\,s^{-1})}$)} \\
        & & \colhead{XRB} & \colhead{0.5-8 keV} \\
        (1) & (2) & (3) & (4)
        }
\startdata
    10045  & 1.404\pm0.171 & 39.85_{-0.70}^{+0.26} & 42.94_{-0.04}^{+0.04} \\
    104881 & 0.139\pm0.040 & 38.65_{-0.45}^{+0.22} & 40.85_{-0.26}^{+0.13} \\
    124477 & 0.007\pm0.000 & 38.14_{-0.17}^{+0.17} & 38.70^{+0.73} \\
    124554 & 0.016\pm0.000 & 38.09_{-0.36}^{+0.19} & \leq39.17             \\
    15235  & 3.578\pm0.218 & 39.97_{-0.67}^{+0.25} & 40.53_{-0.50}^{+0.36} \\
    152627^\dagger & 1.165\pm0.216 & 37.52_{-0.65}^{+0.25} & 42.31_{-0.58}^{+0.24} \\
    156688 & 0.089\pm0.034 & 37.97_{-1.00}^{+0.28} & 39.06^{+0.55} \\
    51928  & 0.253\pm0.031 & 38.68_{-0.46}^{+0.22} & 38.87^{+0.63} \\
    57867  & 0.011\pm0.000 & 37.43_{-0.84}^{+0.27} & \leq39.19             \\
    61072  & 0.129\pm0.002 & 38.69_{-0.72}^{+0.26} & \leq40.35             \\
    67333  & 0.001\pm0.000 & 36.99_{-0.29}^{+0.17} & \leq39.12             \\
    86652  & 0.126\pm0.002 & 38.77^{+0.31}         & 39.28^{+1.04}         \\
    88260  & 0.024\pm0.004 & 37.82_{-0.39}^{+0.20} & \leq39.23             \\
    97904  & 0.066\pm0.002 & 38.91_{-0.73}^{+0.26} & \leq40.12             \\
\enddata
\tablecomments{Col. (2) gives the star formation rates calculated from H$\alpha$ flux (or FUV flux for NSA 152627, which is marked with a dagger). Col. (3) gives the logarithm of the expected luminosities from X-Ray binaries. Col. (4) gives the logarithm of the observed signal in the equivalent X-ray band for comparison. All errors shown are 3-$\sigma$, and upper limits are denoted with $\leq$.}
\end{deluxetable}

\subsection{FeX Emission}
From our analysis, we detect [\ion{Fe}{10}] emission in two of our objects: NSA 10045 and 15235. We calculate fluxes of $11.8\pm3.3$ and $25.3\pm4.4\times10^{-17} erg/s/cm^2$, respectively.

\section{Discussion} \label{sec:discussion}
Four of the fourteen objects in our sample were detected in the broad X-ray band, with two further marginal detections. All are consistent with the respective galaxy nucleus.
For the remaining eight galaxies, we obtain only upper limits on the X-ray luminosity ranging from $10^{39}$ to $2\times10^{40}$ erg/s.
These upper limits correspond to $\sim10^{-5}-10^{-1}\;\rm{L_{Edd}}$ for a $10^{5}\,\rm{M_{\odot}}$ black hole.

\subsection{Origin of the X-ray emission} 
A challenge for X-ray studies of low-mass galaxies is distinguishing between emission from an AGN and X-ray binary.
We compute the expected luminosity from X-ray binaries for each object based on the SFR and stellar mass contained within the \textit{Chandra} PSF.
Of the four galaxies that are detected, all have luminosities exceeding the predicted XRB luminosity.
This suggests that AGNs reside in centers of at least those four galaxies, since their X-ray luminosities likely cannot be attributed to XRBs alone.

However, we think it is unlikely that the remaining sources are solely explained by XRBs.
For all sources, there is likely some contribution to the UV luminosity from the putative AGN, leading to an overestimated SFR (and thus XRB luminosity).
Additionally, the superior angular resolution of \textit{Chandra} allows us to isolate detected X-rays to the nucleus.
We find no off-nuclear X-ray sources detected in our study; all sources were consistent with their optical centers.
Since XRBs are not preferentially found in galaxy nuclei, it would be extremely unusual for these X-ray sources to be nuclear XRBs. 

While we cannot rule out an XRB origin for some of these galaxies, the combination of nuclear variability and a nuclear X-ray point source leads us to conclude that most detections are due to accretion onto a central massive BH.


\subsection{Comparison of AGN Detection Methods} \label{subsec:comparison}

All fourteen galaxies in our sample exhibited optical variability consistent with an AGN.
Four were detected in the broad X-ray band with two additional marginal detections, all having X-ray luminosities greater than expected from XRBs.
However, as discussed above, the fact that the X-ray sources are nuclear and/or point-like is inconsistent with most being XRBs.

\begin{figure}[htb]
    \centering
    \includegraphics[width=0.8\columnwidth]{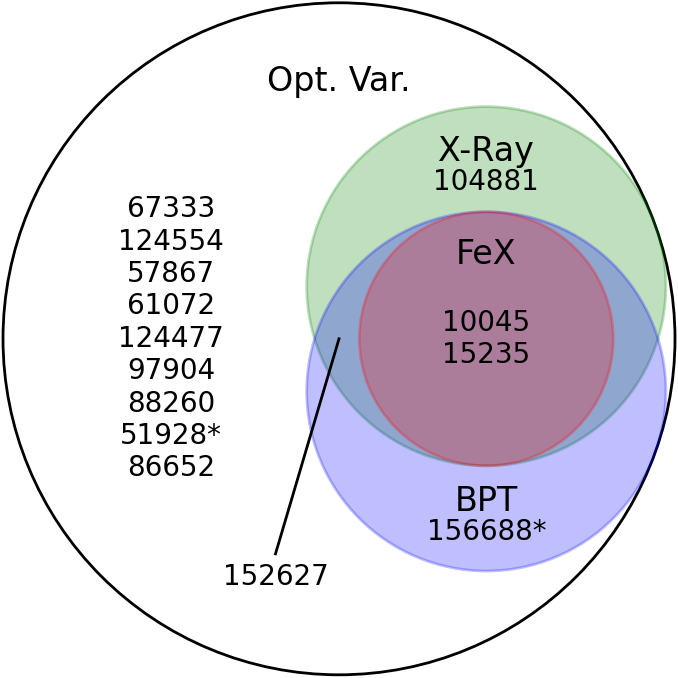}
    \caption{A Venn diagram of the galaxies in our sample based on AGN detection per method: optical variability (the entire sample), X-ray detection (green), emission line diagnostics (BPT; light purple), or the \ion{Fe}{10} coronal line (red). Galaxies that were marginally detected in the X-ray are marked with an asterisk.}
    \label{fig:venn}
\end{figure}

Three of the four X-ray detected objects have emission lines in the AGN or composite region of a BPT diagram based on their SDSS spectroscopy.
Two galaxies, NSA 10045 and 15235, have variability, spectroscopic signatures, and X-ray luminosities significantly higher than expected from XRBs.
A visual breakdown of these results is shown in Figure \ref{fig:venn}.
This outcome can be characterized by discussing the concepts of "pure" versus "complete" samples.
In our study, variability selection provides a more complete sample since it finds AGN candidates that would be missed by other selection techniques, but other methods such as emission line diagnostics and \ion{Fe}{10} emission create a more pure sample since they more exclusive and less likely to containing false negatives.
The lack of substantial overlap in these selection techniques leaves us with some questions: If the variability is due to accretion onto a BH, why are some objects not selected by the BPT diagram or X-ray analysis?


Optical spectroscopic selection has been shown to be incomplete at low galaxy masses. Low metallicity and dilution of the AGN signal by star formation can shift objects out of the AGN region of the BPT diagram \citep{2006MNRAS.372..961K,2015ApJ...811...26T}.
In \citet{Cann_2019}, they use modeling to show that the standard emission line diagnostics could be less reliable for lower mass systems, especially for BHs less massive than about $10^4 M_\odot$.

Our findings are consistent with these results as only four of the ten galaxies plotted on the BPT diagram were categorized as AGNs, despite their optical variability.
Of particular note is galaxy NSA 104881, which was X-ray loud with $L_X\sim 10^{41} \;\rm{erg\;s^{-1}}$, but fell in the star-forming region of the BPT.

\begin{figure}[htb]
    \centering
    \includegraphics[width=0.8\columnwidth]{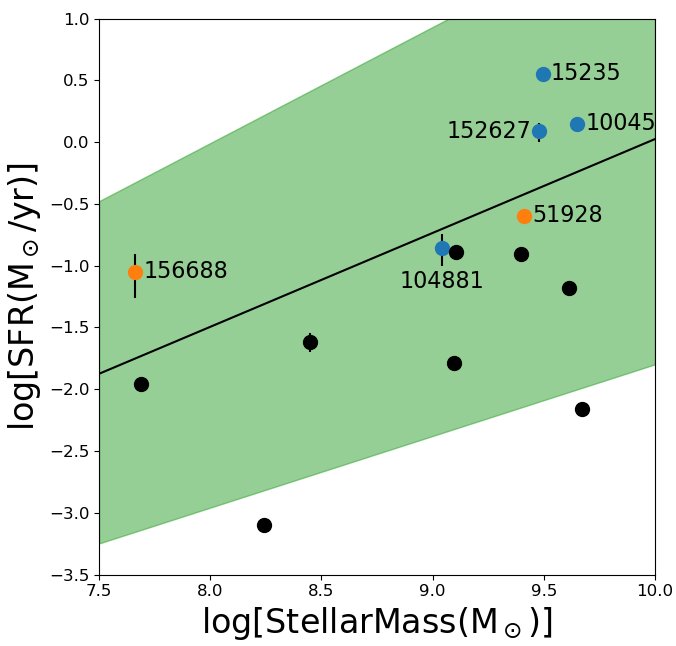}
    \caption{A comparison of the star formation rates calculated with the relation from \citet{2019ApJS..243....3L} versus stellar mass with 3-$\sigma$ errors shown. The black line shows the relation given in \citet{2017MNRAS.465.3390A} with the green region encompassing the error and variation in redshift. The blue points represent objects that were detected in the broad X-ray band, orange points represent marginal detections, and black points were not detected.}
    \label{fig:SFR}
\end{figure}

To quantitatively explore the possibility of star formation dilution, we plot our objects on the star forming main sequence \citep{2017MNRAS.465.3390A}.
The main sequence depends only on galaxy stellar mass and redshift:

\begin{equation}
    \log{\left(\frac{SFR}{M_\odot/yr}\right)} = A+B\log{\left(\frac{M_*}{M_\odot}\right)}+C\log{\left(1+z\right)}
\end{equation}
with $A=-7.6, B=0.76\pm0.06$, and $C=2.95\pm0.33$.
We use this relation to compare against the star formation rates previously calculated in Section \ref{subsec:XRL}, shown in Figure \ref{fig:SFR}.
Many of the galaxies in our sample are consistent with being on the star forming main sequence, showing that this may play a role in some objects being missed by the BPT diagram.
However, this is likely not the only factor at play: the BPT star-forming galaxies the were detected and marginally detected in the X-ray (NSA 104881 and NSA 51928) actually fall slightly below this relation.

We find one object (NSA 51928) that is selected with variability and marginally detected in the X-ray, but whose observed luminosities were not significantly greater than their expected XRB luminosities.
It is possible that X-ray emission could be suppressed in low-mass AGN;
\citet{2012ApJ...761...73D} finds that IMBHs have suppressed X-ray luminosities relative to the UV emission.
They see a trend toward flat spectral indexes $\Gamma$, implying the presence of absorption.
Some, however, are found to have typical gamma values, but low values for $\alpha_{OX}$, the X-ray-to-optical spectral slope, raising the possibility of intrinsically weak X-ray sources.
\citet{2020ApJ...900..141P} also finds a distribution of $\Delta\alpha_{OX}$ that suggests the existence of X-ray weak AGNs, representing $\sim$5\% of the population.
This weakness could be characterized by analyzing each object's X-ray variability, or by searching their UV or X-ray spectra for obscuration signatures.
Follow-up work with UV or X-ray spectroscopy could determine whether obscuration plays a role in this systems.

We must also consider the possibility that the optical variability is not associated with an AGN.
Recall that these galaxies were selected in \citet{2020ApJ...896...10B} for their AGN-like optical variability.
More specifically, these objects were chosen based on $\sigma_{var}$, $\sigma_{QSO}$ and $\sigma_{notQSO}$ from the software QSO\_fit \citep{2011AJ....141...93B}.
These respectively represent the significance that the object is variable, that the object's variability fits a damped random walk, and that the object is not a quasi-stellar object (QSO), but varies in a time-independent Gaussian manner.
Objects that were selected were done so on the basis that $\sigma_{var}>2$, $\sigma_{QSO}>2$ and $\sigma_{QSO}\gtrsim \sigma_{notQSO}$.
Any objects with a burst-like light curve (a rise and fall) were also removed from the sample to eliminate possible contamination from supernovae. 
There are some examples of objects displaying AGN-like variability due to non-AGN processes. The extremely metal poor blue compact emission line galaxy PHL 293B showed variability consistent with a damped random walk for almost two decades \citep{2020ApJ...894L...5B}.
This object was suspected to be an AGN, however, it showed fading broad emission lines with a P Cygni profile and blueshifted absorption lines. \citep{2020ApJ...894L...5B} concluded that this was likely a long-lived, unusual Type II supernova.
While we cannot strictly rule out the possibility of contamination from non-stellar processes, PHL 293B was an extremely rare case and none of the objects in our sample show spectral features like P Cygni profiles or blueshifted absorption.
We consider it unlikely that our sample is significantly contaminated by non-AGN variability. 

Finally, we consider the interesting possibility that the previously detected AGN-variability was associated with a short-lived accretion event like a tidal disruption event.
The PTF data used for constructing light curves was taken between 2009 and 2017.
Our followup \textit{Chandra} X-ray observations were taken in 2020 and 2021.
It is possible that the accretion event leading to the variability has faded in the interim.
Some works have speculated that TDEs could power the entire observed population of AGN in dwarf galaxies \citep{2019MNRAS.483.1957Z}.
X-ray observations that are contemporaneous with observed variability could help shed light on the nature of accretion processes occurring in dwarf galaxies.  

\subsection{Black Hole Mass and Eddington Luminosity} \label{subsec:Eddington}

Here we consider the accretion properties of the X-ray detected AGN in our sample.
We compute or estimate BH masses and Eddington ratios.
The Eddington luminosity is a rough upper limit for energy production in an accreting system where the gravitational force of the black hole on in-falling matter is balanced by the radiation pressure from its emission.
This is related to the central BH's mass:

\begin{equation}
    L_{Edd}\coloneqq\frac{4\pi Gcm_p}{\sigma_T}M_{BH}\approx1.257\times10^{38} \left(\frac{M_{BH}}{M_\odot}\right)\:\rm{erg/s}
\end{equation}

The Eddington ratio can then be thought of as a measure of efficiency for an accreting black hole, the ratio of its bolometric luminosity to the theoretical limit: $k_{Edd}=L_{bol}/L_{Edd}$.
For the sake of this calculation, we use the simple bolometric correction to hard X-rays given in \citet{2004MNRAS.351..169M} $L_{bol}/L_{2-10keV}=10$.
 
Two objects in the sample (NSA 10045 and 15235) have broad H$\alpha$ emission which we can use to estimate BH mass \citep{2005ApJ...630..122G}.
NSA 15235 was previously analyzed in \cite{2017ApJ...836...20B}, and found to have a BH mass of $1.6\times10^{5}\;M_{\odot}$ and an Eddington fraction of $0.3\%$.
NSA 10045 has a BH mass of $1\times10^{6}\;M_{\odot}$ and an Eddington fraction of $33\%$.

To estimate the mass of the remaining central BHs, we use the relation between BH mass and stellar mass found in \citet{2015ApJ...813...82R}:

\begin{equation}
    \log{(M_{BH}/M_\odot)} = \alpha + \beta \log{(M_*/10^{11}M_\odot)}
\end{equation}

where $\alpha = 7.45\pm0.08, \beta = 1.05\pm0.11$.

Assuming this sample follows the BH mass-stellar mass scaling relation, we find BH masses ranging from $10^{4} - 10^{6}\;M_{\odot}$.
For the X-ray detected objects, Eddington fractions from $\sim10^{-5} - 0.3$ (or from $0.001\%$ to $30\%$).
The median Eddington fraction is 0.003 ($0.3\%$).
This is about an order of magnitude lower than the median Eddington fraction for broad line AGN in dwarf galaxies \citep{2017ApJ...836...20B}.
This suggests that variability selection may identify lower accretion rate AGN in dwarf galaxies than searches using broad emission lines.
For the non-detected objects, the upper limits on the Eddington fractions range from $10^{-4}$ to $2\times10^{-1}$.
These results are summarized in Table \ref{table:Eddington}.

\begin{deluxetable*}{C C C C C}[htb]
\tablecolumns{5}
\tablecaption{\label{table:Eddington}}
\tablehead{\colhead{NSA ID} & \colhead{BH Mass} & \colhead{$L_{Edd}$} & \colhead{$L_{2-10 keV}$} & \colhead{$k_{Edd}$} \\
    & \colhead{$\mathrm{log(M_\odot\:yr^{-1})}$} & \colhead{$\mathrm{log(erg\:s^{-1})}$} & \colhead{$\mathrm{log(erg\:s^{-1})}$} & \\
    (1) & (2) & (3) & (4) & (5)
    }
\startdata
    10045  & 6.00^\dagger\pm0.90 & 44.10\pm0.90 & 42.63_{-0.10}^{+0.08} & 3.39\times10^{-1}  \\
    104881 & 5.39\pm0.69 & 43.49\pm0.69 & 40.76_{-0.40}^{+0.21} & 1.86\times10^{-2}  \\
    124477 & 6.05\pm0.50 & 44.15\pm0.50 & 38.40^{+1.08}         & 1.76\times10^{-5}  \\
    124554 & 5.45\pm0.67 & 43.55\pm0.67 & \leq39.31             & \leq2.72\times10^{-3} \\
    15235  & 5.29^\dagger\pm0.90 & 43.39\pm0.90 & 39.74^{+0.98}         & 2.25\times10^{-3} \\ 
    152627 & 5.85\pm0.56 & 43.95\pm0.56 & 41.59^{+0.59}         & 4.33\times10^{-2} \\
    156688 & 3.95\pm1.13 & 42.05\pm1.13 & 38.65^{+0.96}         & 4.02\times10^{-3} \\
    51928  & 5.78\pm0.58 & 43.88\pm0.58 & 39.12^{+0.58}         & 1.72\times10^{-4}  \\
    57867  & 3.97\pm1.12 & 42.07\pm1.12 & \leq39.33             & \leq 2.37\times10^{-1}\\
    61072  & 5.46\pm0.67 & 43.56\pm0.67 & \leq40.80             & \leq8.03\times10^{-2} \\
    67333  & 4.55\pm0.94 & 42.65\pm0.94 & \leq39.32             & \leq4.07\times10^{-2} \\
    86652  & 5.77\pm0.58 & 43.87\pm0.58 & \leq40.63             & \leq2.20\times10^{-2} \\
    88260  & 4.77\pm0.88 & 42.87\pm0.88 & 38.32^{+1.17}         & 2.85\times10^{-4} \\
    97904  & 5.99\pm0.52 & 44.09\pm0.52 & \leq40.40             & \leq6.72\times10^{-3} \\
\enddata
\tablecomments{A table showing rough estimates for BH masses and Eddington ratios. Black hole masses marked with a dagger were calculated broad H$\alpha$ emission, while the others were calculated using the relation given in \citet{2015ApJ...813...82R}. The Eddington ratio was estimated using the simple bolometric correction $L_{bol}/L_{2-10 keV}=10$.Upper limit values are denoted with $\leq$.}
\end{deluxetable*}

\section{Conclusion} \label{sec:conclusion}
We analyze \textit{Chandra} X-ray observations of fourteen nearby ($z\lesssim0.044$) low-mass ($\log{M_*}\lesssim9.7$) galaxies from the NASA-Sloan Atlas, selected for their optical variability in \citet{2020ApJ...896...10B}.
\begin{itemize}
    \item 4 of the 14 objects (29\%) in our sample of variability-selected low-mass AGN were detected in the 0.5-7 keV X-ray band, with two further marginal detections.
    \item Of the 4 X-ray detected, all had luminosities exceeding that anticipated from XRBs by multiple $\sigma$, which can be interpreted as evidence for AGN activity.
    \item One of the galaxies with an X-ray detection and one marginally detected galaxy fall in the star forming region of the BPT diagram. This work confirms that variability can find AGN in dwarf galaxies that are missed by optical spectroscopy. 
    \item BPT analysis, which plots the ratios of specific emission lines with empirical cuts that categorize galaxies, classified four of our objects (NSA 10045, 15235, 152627, and 156688) as AGN candidates. This method, however, is known to undercount low-mass and low-metallicity systems.
    \item Based on scaling relations between BH mass and stellar mass, we find Eddington fractions ranging from $10^{-5} - 0.3$ for the X-ray detected objects in our sample. These are systematically lower Eddington fraction than broad-line AGN in dwarf galaxies, implying that variability selection may find lower accretion rate objects. 
\end{itemize}

With the Vera C. Rubin Observatory Legacy Survey of Space and Time (LSST; \citealt{2019ApJ...873..111I}) on the horizon, we can expect an abundance of new variability-selected candidate AGN in low-mass galaxies. Contamination from stellar processes may also be higher since imaging will be deeper than previous time domain surveys. In order to better understand the source of variability in these future candidates (i.e., AGN versus TDE versus stellar processes), it will be useful to combine variability observations from LSST with simultaneous X-ray observations.  



\section{Acknowledgements} \label{sec:acknowledgements}
Alexander Messick thanks the LSSTC Data Science Fellowship Program, which is funded by LSSTC, NSF Cybertraining Grant \#1829740, the Brinson Foundation, and the Moore Foundation; their participation in the program has benefited this work.

Support for this work was provided by the National Aeronautics and Space Administration through Chandra Award Number GO1-22100X issued by the Chandra X-ray Center, which is operated by the Smithsonian Astrophysical Observatory for and on behalf of the National Aeronautics Space Administration under contract NAS8-03060. The scientific results reported in this article are based to a significant degree on observations made by the Chandra X-ray Observatory and data obtained from the Chandra Data Archive. This research has made use of software provided by the Chandra X-ray Center (CXC) in the application package CIAO.

Funding for SDSS-III has been provided by the Alfred P. Sloan Foundation, the Participating Institutions, the National Science Foundation, and the U.S. Department of Energy Office of Science. The SDSS-III web site is http://www.sdss3.org/. SDSS-III is managed by the Astrophysical Research Consortium for the Participating Institutions of the SDSS-III Collaboration including the University of Arizona, the Brazilian Participation Group, Brookhaven National Laboratory, Carnegie Mellon University, University of Florida, the French Participation Group, the German Participation Group, Harvard University, the Instituto de Astrofisica de Canarias, the Michigan State/Notre Dame/JINA Participation Group, Johns Hopkins University, Lawrence Berkeley National Laboratory, Max Planck Institute for Astrophysics, Max Planck Institute for Extraterrestrial Physics, New Mexico State University, New York University, Ohio State University, Pennsylvania State University, University of Portsmouth, Princeton University, the Spanish Participation Group, University of Tokyo, University of Utah, Vanderbilt University, University of Virginia, University of Washington, and Yale University. 

This publication makes use of data products from the Wide-field Infrared Survey Explorer, which is a joint project of the University of California, Los Angeles, and the Jet Propulsion Laboratory/California Institute of Technology, funded by the National Aeronautics and Space Administration.

Some of the data presented in this paper were obtained from the Infrared Science Archive (IRSA) at IPAC, which is operated by the California Institute of Technology under contract with the National Aeronautics and Space Administration. The specific observations analyzed can be accessed via \dataset[doi:10.26131/IRSA1]{https://doi.org/10.26131/IRSA1}. Some of the data presented in this paper were obtained from the Mikulski Archive for Space Telescopes (MAST) at the Space Telescope Science Institute. The specific observations analyzed can be accessed via \dataset[doi:10.17909/T9H59D]{https://doi.org/10.17909/T9H59D}.

\bibliography{AGN_Follow_Up}{}
\bibliographystyle{aasjournal}

\end{document}